\documentclass[preprint,12pt]{elsarticle}

\usepackage{amsthm}
\usepackage{amsmath}
\usepackage{amssymb}
\usepackage{amstext}
\usepackage{bm}
\usepackage{graphicx}
\usepackage{epstopdf}
\usepackage{subcaption}

 \newcommand{\bl}{\big<}
  \newcommand{\bg}{\big>}

\newcommand{\ux}{{\bf x}}

\newcommand{\uomega}{{\bf \Omega}}
\newcommand{\unabla}{{\bf \nabla}}

\biboptions{sort&compress}

\journal{ArXiv}

\begin{document}

\begin{frontmatter}

\title{The Nonclassical Diffusion Approximation to the Nonclassical Linear Boltzmann Equation}

\author[ucb]{Richard Vasques\corref{cor1}}
\cortext[cor1]{Corresponding author: richard.vasques@fulbrightmail.org\\
Postal address: 4103 Etcheverry Hall, MC 1730, University of California, Berkeley\\
Berkeley, CA 94720-1730, United States of America}
\address[ucb]{Department of Nuclear Engineering, University of California, Berkeley}

\begin{abstract}

We show that, by correctly selecting the probability distribution function $p(s)$ for a particle's distance-to-collision, the nonclassical diffusion equation can be represented exactly by the nonclassical linear Boltzmann equation for an infinite homogeneous medium. This choice of $p(s)$ preserves the \textit{true} mean-squared free path of the system, which sheds new light on the results obtained in previous work.

\end{abstract}

\end{frontmatter}

\section{Introduction}\label{sec1}
\setcounter{section}{1}
\setcounter{equation}{0} 

A \textit{nonclassical linear Boltzmann equation} has been recently proposed to address particle transport problems in which the particle flux experiences a nonexponential attenuation law \cite{larsen_11, frank_10, vasques_14a}. This nonexponential behavior arises in certain inhomogeneous random media in which the locations of the scattering centers are spatially correlated, such as in a Pebble Bed reactor core \cite{larsen_11, vasques_13, vasques_14b}. 

Independent of these developments, a similar kinetic equation has been
rigorously derived for the periodic Lorentz gas in a series of papers by Golse (cf. \cite{golse_12}), and by Marklof and Str\" ombergsson \cite{marklof_11, marklof_15}. Related work has also been performed by Grosjean \cite{grosjean_51}, considering a generalization of
neutron transport that includes arbitrary path-length distributions.

For the case of monoenergetic particle transport with isotropic scattering, the nonclassical linear Boltzmann equation is written as
\begin{equation}
\label{eq1.1}
\begin{split}
\frac{\partial\psi}{\partial s}(\ux,\uomega,s) &+ \uomega\cdot\unabla \psi(\ux,\uomega,s) + \Sigma_t(s)\psi(\ux,\uomega,s)\\
&= \frac{\delta(s)}{4\pi}\left[ c\int_{4\pi}\int_0^\infty \Sigma_t(s')\psi(\ux,\uomega',s')ds' d\Omega' + Q(\ux) \right],
\end{split}
\end{equation}
where $\psi$ is the nonclassical angular flux, $c$ is the scattering ratio (probability of scattering), and $Q(\bf x)$ is a source. Here, the total cross section $\Sigma_t$ is a function of the path length $s$ (distance traveled by the particle since its previous interaction), such that the path length distribution  
\begin{equation}\label{eq1.2}
p(s) = \Sigma_t(s)e^{-\int_0^s \Sigma_t(s')ds'}
\end{equation}
does not have to be exponential. If $p(s)$ is exponential, Eq.\ (\ref{eq1.1}) reduces to the classical linear Boltzmann equation
\begin{equation}
\label{eq1.3}
\uomega\cdot\unabla \psi(\ux,\uomega) + \Sigma_t \psi(\ux,\uomega) = \frac{\Sigma_s}{4\pi}\int_{4\pi}\psi(\ux,\uomega')d\Omega'+ \frac{Q(\ux)}{4\pi}
\end{equation}
for the classical angular flux 
\begin{equation}
\label{eq1.4}
\psi(\ux,\uomega) = \int_0^\infty \psi(\ux,\uomega,s)ds.
\end{equation} 
It has been shown \cite{siap15} that, by selecting $\Sigma_t(s)$ in a proper way, Eq.\ \eqref{eq1.1} can be converted to an integral equation for the scalar flux
\begin{equation}
\label{eq1.5}
\phi_0(\ux) = \int_{4\pi} \psi(\ux,\uomega)d\Omega
\end{equation}
that is identical to the integral equation that can be constructed for certain diffusion-based approximations to Eq.\ \eqref{eq1.3} in the hierarchy of the $SP_N$ equations \cite{mcclarren_11}.

The work in this paper shows that this is also the case for the \textit{nonclassical diffusion equation} \cite{larsen_11}
\begin{equation}
\label{eq1.6}
-\frac{\bl s^2\bg}{6\bl s\bg} \nabla^2 \phi_0(\ux) + \frac{1-c}{\bl s\bg} \phi_0(\ux) = Q(\ux),
\end{equation}
which is an asymptotic approximation of Eq.\ (\ref{eq1.1}) when $\bl s^2\bg < \infty$.
Here,
\begin{equation}
\label{eq1.7}
\bl s\bg = \int_0^{\infty} sp(s)ds \,\,\,\, \text{and} \,\,\,\, \bl s^2\bg = \int_0^{\infty} s^2p(s)ds.
\end{equation}
Specifically, we find $p(s)$ and the corresponding $\Sigma_t(s)$ such that the integral equation for Eq.\ (\ref{eq1.6}) is identical to the integral equation for Eq.\ (\ref{eq1.1}). We also show that the second moment of $p(s)$ for nonclassical diffusion preserves the \textit{true} mean-squared free path $\bl s^2\bg$, which gives a new insight on the results obtained in \cite{siap15} for the classical diffusion approximations.

The remainder of this paper is organized as follows. In section \ref{sec2} we convert Eq.\ \eqref{eq1.1} to an integral equation for the scalar flux given in Eq.\ \eqref{eq1.5}. In section \ref{sec3} we convert Eq.\ \eqref{eq1.6} to an integral equation for the scalar flux, and find the correct choice of $p(s)$ such that the integral equation obtained in section \ref{sec2} is identical to this nonclassical diffusion integral equation. We also present a numerical example illustrating the differences on $p(s)$ and $\Sigma_t(s)$ for classical transport, classical diffusion, and nonclassical diffusion. The paper concludes with a discussion in section \ref{sec4}.

\section{Integral Equation Formulation}\label{sec2}
\setcounter{section}{2}
\setcounter{equation}{0} 

Let $S(\ux)$ be given by
\begin{subequations}
\begin{align}\label{eq2.1a}
S(\ux) &= c\int_{4\pi}\int_0^\infty \Sigma_t(s')\psi(\ux,\uomega',s')ds'd\Omega'+Q(\ux) \\
&= cf(\ux) + Q(\ux), \nonumber
\end{align}
where 
\begin{align}
f(\ux) &= \int_0^\infty \Sigma_t(s')\phi_0(\ux,s')ds' = \text{collision-rate density}\label{eq2.1b}
\end{align}
and
\begin{align}
\phi_0(\ux,s) &= \int_{4\pi} \psi(\ux,\uomega,s)d\Omega = \text{nonclassical scalar flux}.\label{eq2.1c}
\end{align}
\end{subequations}
We can now write Eq.\ \eqref{eq1.1} as the initial value problem
\begin{subequations}
\label{eq2.2}
\begin{align}
&\frac{\partial\psi}{\partial s}(\ux,\uomega,s) + \uomega\cdot\unabla\psi(\ux,\uomega,s) + \Sigma_t(s)\psi(\ux,\uomega,s) = 0, \;\; 0 < s ,\\
& \psi(\ux,\uomega,0) = \frac{S(\ux)}{4\pi}.
\end{align}
\end{subequations}
Following the steps presented in \cite{larsen_11} and \cite{siap15}, we: (i) use the method of characteristics to calculate the solution of Eqs.\ \eqref{eq2.2}; (ii) operate on this solution by $\int_{4\pi}\int_0^\infty \Sigma_t(s) (\cdot )dsd\Omega$; and (iii) perform the change of spatial variables from the 3-D spherical $(\uomega,s)$ to the 3-D Cartesian $\ux'$ defined by $\ux'= \ux-s\uomega$. This yields
\begin{equation}
\label{eq2.3}
f(\ux) = \int\int\int S(\ux') \frac{p(|\ux'-\ux|)}{4\pi |\ux'-\ux|^2} dV',
\end{equation}
where $p(|\ux'-\ux|)$ and $S(\ux)$ are given by Eqs.\ \eqref{eq1.2} and \eqref{eq2.1a}, respectively.

\section{Nonclassical Diffusion}\label{sec3}
\setcounter{section}{3}
\setcounter{equation}{0} 

The nonclassical diffusion formulation presented in Eq.\ \eqref{eq1.6} is an asymptotic approximation of Eq.\ \eqref{eq1.1} in the case of $\bl s^2\bg<\infty$. In this formulation, $\bl s \bg$ and $\bl s^2\bg$ as given by Eq.\ \eqref{eq1.7} represent the first and second moments of the \textit{true} path length distribution $p(s)$.
 If $p(s)$ is exponential, then $\bl s\bg = 1/\Sigma_t$, $\bl s^2\bg = 2/\Sigma_t^2$, and Eq.\ \eqref{eq1.6} reduces to the classical diffusion equation
\begin{equation}
\label{eq3.1}
-\frac{1}{3\Sigma_t} \nabla^2 \phi_0(\ux) + (1-c)\Sigma_t \phi_0(\ux) = Q(\ux).
\end{equation}

For the general case in which $p(s)$ is \textit{not} assumed to be an exponential, we define $S(\ux) = c\bl s\bg^{-1}\phi_0(\ux)+Q(\ux)$ and rewrite 
Eq.\ \eqref{eq1.6} as:
\begin{subequations}\label{eq3.2}
\begin{align}
-\nabla^2\phi_0(\ux) + \lambda^2 \phi_0(\ux) = \lambda^2 \bl s \bg S(\ux),\label{eq3.2a}
\end{align}
where
\begin{align}
\lambda^2=\frac{6}{\bl s^2 \bg}.
\end{align}
\end{subequations}
The Green's function for the operator on the left hand side of Eq.\ \eqref{eq3.2a} is:
\begin{equation}
\label{eq3.3}
G(|\ux-\ux'|) = \frac{e^{-\lambda |\ux-\ux'|}}{4\pi |\ux-\ux'|};
\end{equation}
therefore, we can transform Eq.\ \eqref{eq3.2a} into an integral equation for $\phi_0(\bf x)$ by taking
\begin{align}\label{eq3.4}
\phi_0(\ux) &= \int\int\int G(|\ux-\ux'|) \lambda^2\bl s \bg S(\ux') dV' \\
&= \int\int\int \frac{\lambda^2\bl s \bg e^{-\lambda |\ux-\ux'|}}{4\pi  |\ux-\ux'|} S(\ux') dV'  \nonumber \\
&= \int\int\int \frac{\lambda^2 \bl s \bg |\ux-\ux'| e^{-\lambda |\ux-\ux'|}}{4\pi |\ux-\ux'|^2} S(\ux') dV'.\nonumber
\end{align}
Bearing in mind that $\bl s\bg$ represents the mean free path of a particle (i.e. the average distance between collisions), the collision-rate density can be written as $f(\ux)=\bl s\bg^{-1} \phi_0(\ux)$, such that
\begin{equation}
\label{eq3.5}
f(\ux)=\frac{\phi_0}{\bl s \bg} = \int\int\int \frac{\lambda^2 |\ux-\ux'| e^{-\lambda |\ux-\ux'|}}{4\pi |\ux-\ux'|^2} S(\ux') dV'.
\end{equation}
This result agrees with Eq.\ \eqref{eq2.3} iff
\begin{equation}
\label{eq3.6}
p(s) = \lambda^2se^{-\lambda s}=\frac{6se^{-\sqrt{6/<s^2>}s}}{\bl s^2\bg}.
\end{equation}
It is easy to verify that
\begin{equation}\label{eq3.7}
\int_0^\infty p(s)ds = \int_0^\infty \frac{6se^{-\sqrt{6/<s^2>}s}}{\bl s^2\bg} ds = 1,
\end{equation}
which shows that Eq.\ \eqref{eq3.6} is a distribution function. Furthermore, $\Sigma_t(s)$ is given by
\begin{equation}
\label{eq3.8}
\Sigma_t(s) = \frac{p(s)}{\int_s^\infty p(s')ds'} = \frac{\lambda^2 s}{1+\lambda s}.
\end{equation}
This shows that the nonclassical transport equation reproduces the nonclassical diffusion approximation given by Eq.\ \eqref{eq1.6} if $p(s)$ and $\Sigma_t(s)$ are defined by Eqs.\ \eqref{eq3.6} and \eqref{eq3.8}. Moreover, if $p(s)$ is exponential, this results agrees with the $p(s)$ and $\Sigma_t(s)$ obtained for the classical diffusion equation in \cite{siap15}.  

We also point out that
\begin{subequations}\label{eq3.9}
\begin{align}
\int_0^\infty sp(s)ds &=\int_0^\infty \frac{6s^2e^{-\sqrt{6/<s^2>}s}}{\bl s^2\bg} ds = \frac{\sqrt{6\bl s^2\bg}}{3},\label{eq3.9a}\\
\int_0^\infty s^2p(s)ds &=\int_0^\infty \frac{6s^3e^{-\sqrt{6/<s^2>}s}}{\bl s^2\bg} ds = \bl s^2\bg\label{eq3.9b}.
\end{align}
\end{subequations}
The first moment of $p(s)$ only approximates the mean free path $\bl s \bg$, as can be seen in Eq.\ \eqref{eq3.9a}. However, Eq.\ \eqref{eq3.9b} shows that the \textit{true} mean-squared free path $\bl s^2\bg$ is preserved. This sheds new light on the results presented in \cite{siap15}, where the second moments of the path length distributions obtained for all (classical) diffusion approximations give the exact (classical) transport value $2/\Sigma_t^2$.

\begin{figure}
    \centering
        \includegraphics[scale=0.3]{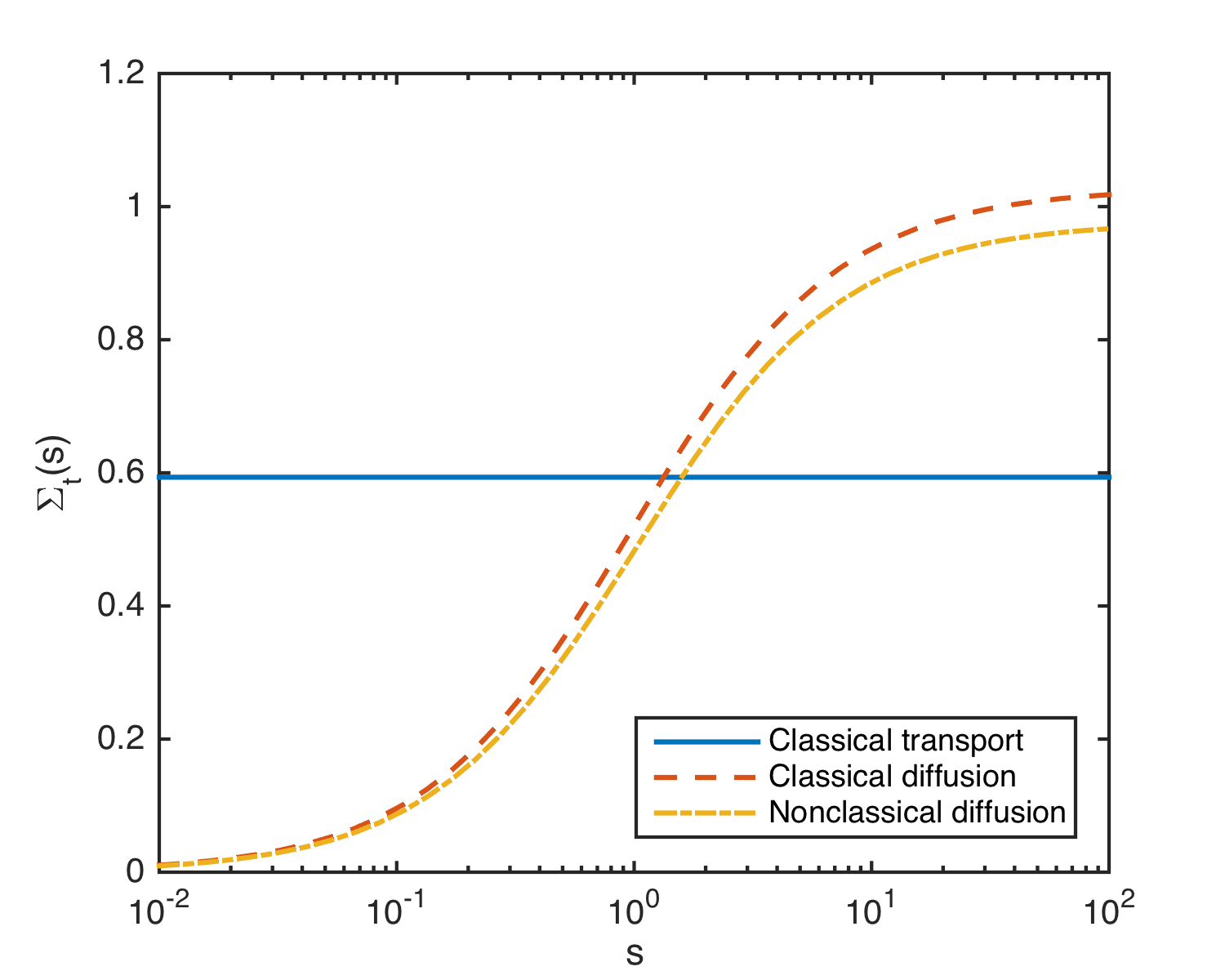}
        \caption{Total cross section as a function of $s$}
        \label{fig1}
\end{figure}

\begin{figure}
    \centering
        \includegraphics[scale=0.3]{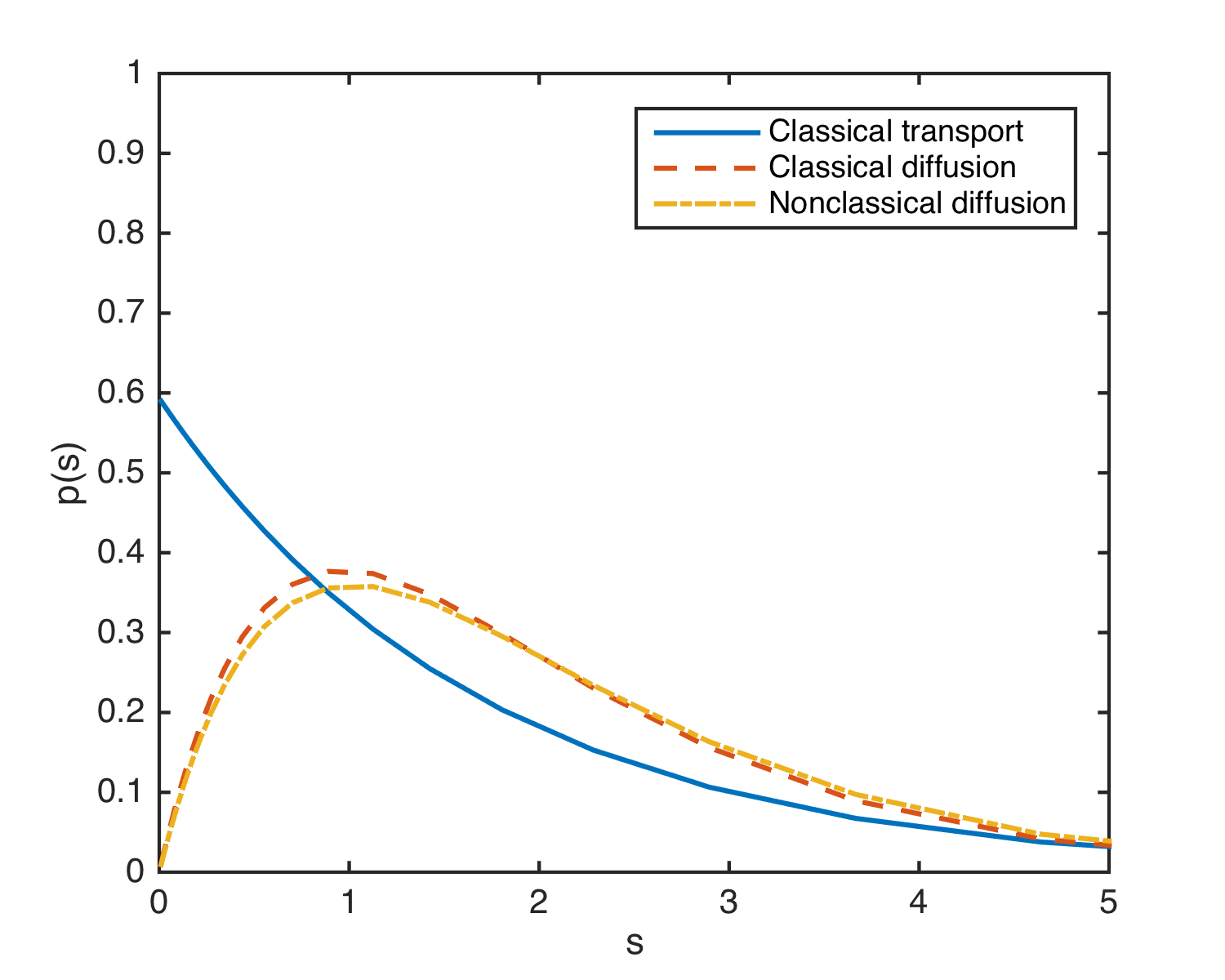}
        \caption{Path length distribution function $p(s)$}
        \label{fig2}
\end{figure}

Figures \ref{fig1} and \ref{fig2} show the functions $\Sigma_t(s)$ and $p(s)$ for the transport of neutrons taking place in the interior of a homogenized 3-D Pebble Bed reactor core system, as described in \cite{vasques_14b}. The system consists of fuel spheres (pebbles) randomly packed in a background void with packing fraction $0.5934$. The parameters for the material of the fuel pebbles are: diameter $d=1$; total cross section $\Sigma_t=1$; and scattering ratio $c=0.99$. As discussed in detail in \cite{vasques_14b}, the \textit{classically homogenized} system (using the Atomic Mix model) has total cross section $\overline \Sigma_t  = 0.5934\Sigma_t = 0.5934$, and its \textit{true} mean-squared free path is numerically calculated to be $\bl s^2 \bg = 6.2898$. We note that nonclassical transport occurs due to the spatial correlations of the fuel pebbles in the system, and therefore $\bl s^2 \bg \neq 2/\overline\Sigma_t^2$.

\section{Discussion}\label{sec4}
\setcounter{section}{4}
\setcounter{equation}{0} 

We have shown that the nonclassical diffusion equation for an infinite homogeneous medium can be represented exactly by the nonclassical linear Boltzmann equation with the correct choice of $\Sigma_t(s)$ and $p(s)$. We derived explicit expressions for these quantities and showed that, while the first moment of the path length distribution $p(s)$ only approximates the true mean free path $\bl s\bg$, its second moment \textit{preserves} the true mean-squared free path $\bl s^2\bg$. This result provides a deeper understanding of the results presented in \cite{siap15} for the second moment of the path-length distributions for classical diffusion approximations.

The work on this paper allows us to construct $p(s)$ and $\Sigma_t(s)$ that yield the correct solution for the nonclassical Boltzmann equation \eqref{eq1.1} in a diffusive system. The only parameter necessary for this is the true mean-squared free path $\bl s^2\bg$. This paves the road to the possibility of using this easy-to-obtain $p(s)$ to approximate the solutions of the nonclassical Boltzmann equation as the system moves away from the diffusive limit. Further work needs to be done to investigate how well such approach would perform; this task, however, must be left for future work.   

\section*{Acknowledgments}

This paper was prepared by Richard Vasques under award number NRC-HQ-84-14-G-0052 from the Nuclear Regulatory Commission. The statements, findings, conclusions, and recommendations are those of the author and do not necessarily reflect the view of the US Nuclear Regulatory Commission.

\pagebreak


\pagebreak

\end{document}